\documentclass[twocolumn,prb]{revtex4}

\usepackage{graphicx}
\usepackage{amsmath}
\usepackage{bm}
\usepackage{comment}
\usepackage{multirow}
\usepackage{color}
\newcommand{\red}{}
\newcommand{\blue}{}

\begin{document}
	\title{Spin-orbit enabled unconventional Stoner magnetism}

\author{Yue Yu$^1$, Tatsuya Shishidou$^1$, Shuntaro Sumita$^{2, 3, 4}$, Michael Weinert$^1$, and Daniel F. Agterberg$^1$}
\affiliation{$^1$Department of Physics, University of Wisconsin--Milwaukee, Milwaukee, Wisconsin 53201, USA}
\affiliation{$^2$Department of Basic Science, The University of Tokyo, Meguro, Tokyo 153-8902, Japan}
\affiliation{$^3$Komaba Institute for Science, The University of Tokyo, Meguro, Tokyo 153-8902, Japan}
\affiliation{$^4$Condensed Matter Theory Laboratory, RIKEN CPR, Wako, Saitama 351-0198, Japan}
\date{\today}

	\begin{abstract}
The Stoner instability remains a cornerstone for understanding metallic ferromagnets. This instability captures the interplay of Coulomb repulsion, Pauli exclusion, and two-fold fermionic spin degeneracy. In materials with spin-orbit coupling, this fermionic spin is generalized to a two-fold degenerate pseudospin {\blue which is typically believed to have symmetry properties as spin}. Here we identify a distinct symmetry of this pseudospin that forbids it to couple to a Zeeman field.
This `spinless' property is required to exist in five non-symmorphic space groups {\blue and has non-trivial implications for superconductivity and magnetism}.  With Coulomb repulsion, Fermi surfaces composed primarily of this spinless pseudospin feature give rise to Stoner instabilities into magnetic states that are qualitatively different than ferromagnets. These spinless-pseudospin ferromagnets break time-reversal symmetry, have a vanishing magnetization, are non-collinear, and exhibit altermagnetic-like energy band spin-splittings. {\blue In superconductors, for all pairing symmetries and field orientations, this spinless pseudospin extinguishes paramagnetic limiting.}  We discuss  applications to {\blue superconducting} UCoGe and \blue{magnetic} NiS$_{2-x}$Se$_x$. 

   	\end{abstract}
	\maketitle

Spin-orbit coupling (SOC) is an essential interaction in quantum materials. It underlies realizations of the Kitaev model \cite{Jackeli:2009}, topological insulators \cite{Schnyder:2008}, electric field control of ferromagnets and antiferromagnets \cite{Baltz:2018}, and high critical fields in Ising superconductors \cite{Wickramaratne:2023}. SOC typically does not alter the symmetry properties of the underlying fermionic Bloch spins. In particular, when time-reversal ($T$) and inversion ($I$) symmetries are present, the two-fold Kramers' degeneracy, or pseudospin, that appears when SOC is present typically behaves as usual spin-1/2 under rotations. An immediate consequence is that a Zeeman magnetic field couples to this pseudospin and can be used to control the quantum states discussed above. Here we find a remarkable {\red scenario, where} the fermionic Bloch pseudospin does not couple to a Zeeman field, no matter which direction this field is applied.

 {\red This `spinless' scenario} is a consequence of two perpendicular glide mirror symmetries, preserving Bloch's momentum on {\red high-symmetry} lines in momentum space. {\red With SOC, the bands remain doubly degenerate on these lines, but the pseudospin excitations preserve both mirror symmetries, preventing coupling to magnetic fields.} The magnetic and {\blue superconducting} response of Fermi surfaces near these momentum lines is consequently dominated by the spinless pseudospin {\red feature}. {\blue The lack of coupling to a Zeeman implies that paramagnetic limiting is suppressed for all superconducting states. } Moreover, when we examine Stoner instabilities \cite{Stoner:1938} of such Fermi surfaces, we find that spinless-pseudospin ferromagnets are energetically much more favorable than usual ferromagnets. 

Unlike usual ferromagnets, these pseudospin ferromagnets {\red preserve all mirror symmetries}, lack net magnetization, are non-collinear in real space and exhibit altermagnetic-like spin-textures in the Brillouin zone (BZ)  \cite{Smejkal:2022}. We find they occur {\blue independently of any microscopic electronic details, such as orbital content or Wyckoff position symmetry,} for the orthorhombic space groups Pbcm, Pbcn, Pbca, Pnma, and the cubic space group Pa$\overline{3}$. {\blue While this may seem a small number of space groups, we note that these five space groups account for 238 out of the 1011 translation invariant magnetic materials in the Magndata database \cite{Gallego:2016}.} {\red Remarkably, for these different five space groups, we find they all exhibit Stoner instabilities into pseudospin ferromagnetic states with shared properties}. These include a non-zero scalar-spin chirality $\chi_{ijk}=\vec{S}_i\cdot(\vec{S}_j\times \vec{S}_k)$ \cite{Wen:1989}, generating $\pm \pi/2$ gauge spin flux in each elementary plaquette and symmetry required Weyl lines in the BZ that imply drumhead surface states appear at mirror-invariant surfaces \cite{Chiu:2014,Chan:2016,Belopolski:2019}. 

Here we examine the application of the spinless pseudospin to {\red superconducting magnetic response and magnetic instabilities. In UCoGe, we perform DFT calculations and reveal vanishing magnetic g-factors for all field orientations, which provides an alternative explanation for the large upper critical field in its superconducting state \cite{huy:2008,aoki:2009}. We analyze Stoner's instability towards unconventional magnetism, providing an itinerant electron mechanism for stabilizing translation invariant non-collinear magnetism as observed in NiS$_{2-x}$Se$_x$ under chemical pressure \cite{Friedemann:2016, Yano:2016}.}

\begin{figure}[h]
	\centering
	\includegraphics[width=7cm]{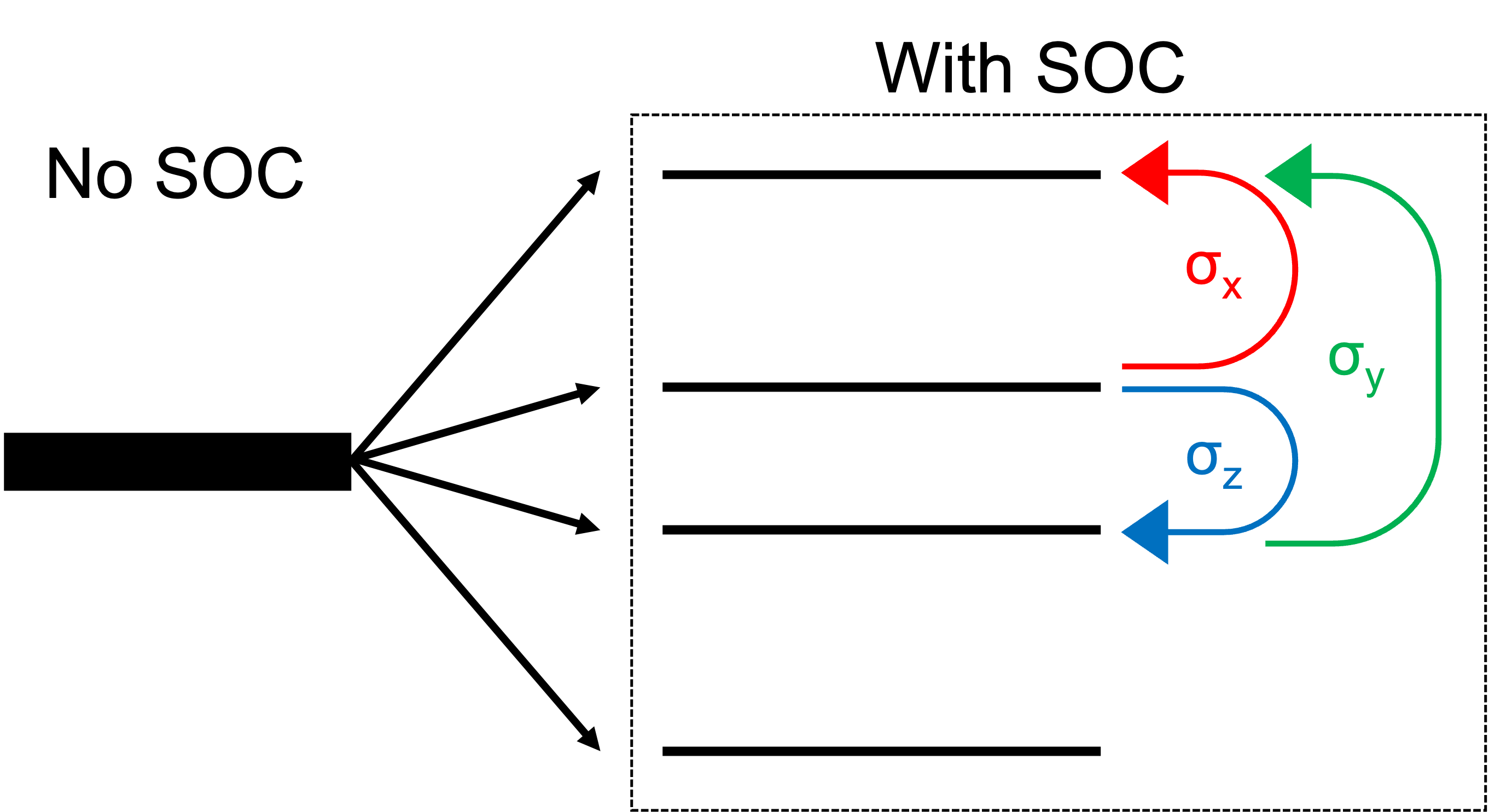}
	\caption{At Brillouin zone edges hosting spinless pseudospin, SOC splits the 8-fold degenerate band into 4 doubly degenerate bands, leaving all spin operators inter-band.}\label{F:Band}
\end{figure}

\vglue 0.1 cm
\noindent{\it Symmetries of pseudospin Pauli Matrices:}  Our key finding is that in the five space groups listed above, SOC {\red makes spin operators inter-band and forbids pseudospins to }couple to a Zeeman field (Fig.\ref{F:Band}). Before discussing our general results, we highlight the symmetry differences between the usual pseudospins ({\red i.e. electron spins}) and the spinless pseudospins found here. 

\begin{table}[ht]
    \centering
    \begin{tabular}{c|c|c}
       Class&$C_{2v}$ Symmetry& $\tilde{\sigma}_i$  \\ \hline
       spinless pseudospin&$A_1$ & $\tilde{\sigma}_x=k_zk_y\sigma_x$\\ &$A_1$&$\tilde{\sigma}_y=k_zk_x\sigma_y$\\ & $A_1$&$\tilde{\sigma_z}=k_xk_y\sigma_z$
    \\ \hline
    usual pseudospin&$B_2$ & $\tilde{\sigma}_x=\sigma_x$\\
    &$B_1$  & $\tilde{\sigma}_y=\sigma_y$\\
    &$A_2$ & $\tilde{\sigma}_z=\sigma_z$\\
    \end{tabular}
    \caption{Symmetry distinction between spinless pseudospin Pauli matrices and usual pseudospin Pauli matrices.  Since all spinless pseudospin components belong to a different symmetry than the usual spin, it cannot couple to a Zeeman field.}
    \label{T:1}
\end{table}

Central to the {\red spinless property} are lines in the BZ that are invariant under the point group $C_{2v}$, which consists of the identity $E$, two mirror symmetries $M_x$ and $M_y$, and their product $C_{2z}$. At each ${\bf k}$ on this line, labeling the Kramers' degenerate states as $|1\rangle$ and  $|2\rangle$, we can define pseudospin Pauli matrices  $\tilde{\sigma}_{x}=|1\rangle \langle 2|+|2\rangle \langle 1|$,  $\tilde{\sigma}_{y}=i|1\rangle \langle 2|-i|2\rangle \langle 1|$, and  $\tilde{\sigma}_{z}=|1\rangle \langle 1|-|2\rangle \langle 2|$. We find that on these $C_{2v}$ invariant lines, {\red instead of} the usual pseudospin, spinless pseudospin can exist. These two {\red scenarios} differ in their symmetry transformation properties under $C_{2v}$. As summarized in Table 1, for usual pseudospin {\red scenario}, the pseudospin Pauli matrices belong to the $A_2$, $B_1$, and $B_2$ representations of the group $C_{2v}$, same as usual spin-1/2 Pauli matrices denoted by $\sigma_i$. For spinless pseudospin {\red scenario}, all pseudospin Pauli matrices $\tilde{\sigma}_i$ belong to the $A_1$ representation, ensuring no coupling to a Zeeman field. Spinless pseudospin Pauli matrices in terms of usual spin-1/2 Pauli matrices necessarily include some momentum-dependent {\red or orbital-dependent} functions. In Table 1, we carry out such a representative expansion valid near the $\Gamma$ point. 

{\red To ensure that pseudospin excitations preserve mirror $M_{x,y}$ symmetries and belong to the $A_1$ representation, the Kramers' degenerate states $|1\rangle$ and $|2\rangle=TI|1\rangle$ need to share the same $M_{x}$ (and $M_{y}$) eigenvalues. This is achievable with non-symmorphic symmetries, for which mirror and inversion operation generically do not commute. Analysis of the momentum-dependent commutator across all space groups reveals that pseudospin must become spinless along specific momentum lines: in Pbcm (SG 57) along $(k_x,\pi,\pi)$; in Pbcn (SG 60) along$(\pi,k_y,\pi)$; in Pnma  (SG 62) along$(\pi,\pi,k_z)$; in Pbca (SG 61) along  $(k_x,\pi,\pi), (\pi,k_y,\pi)$, and  $(\pi,\pi,k_z)$;  and in Pa$\overline{3}$ (SG 205) along $(k_x,\pi,\pi), (\pi,k_y,\pi)$, and $(\pi,\pi,k_z)$. The detailed symmetry analysis is in the supplementary material. Notably, in previous studies without considering SOC\cite{Takahashi:2017,Li:2021,Suh:2023}, these lines exhibit eight-fold band degeneracy (Fig.\ref{F:Band}).   }

\begin{figure}[h]
		\centering
		\includegraphics[width=7cm]{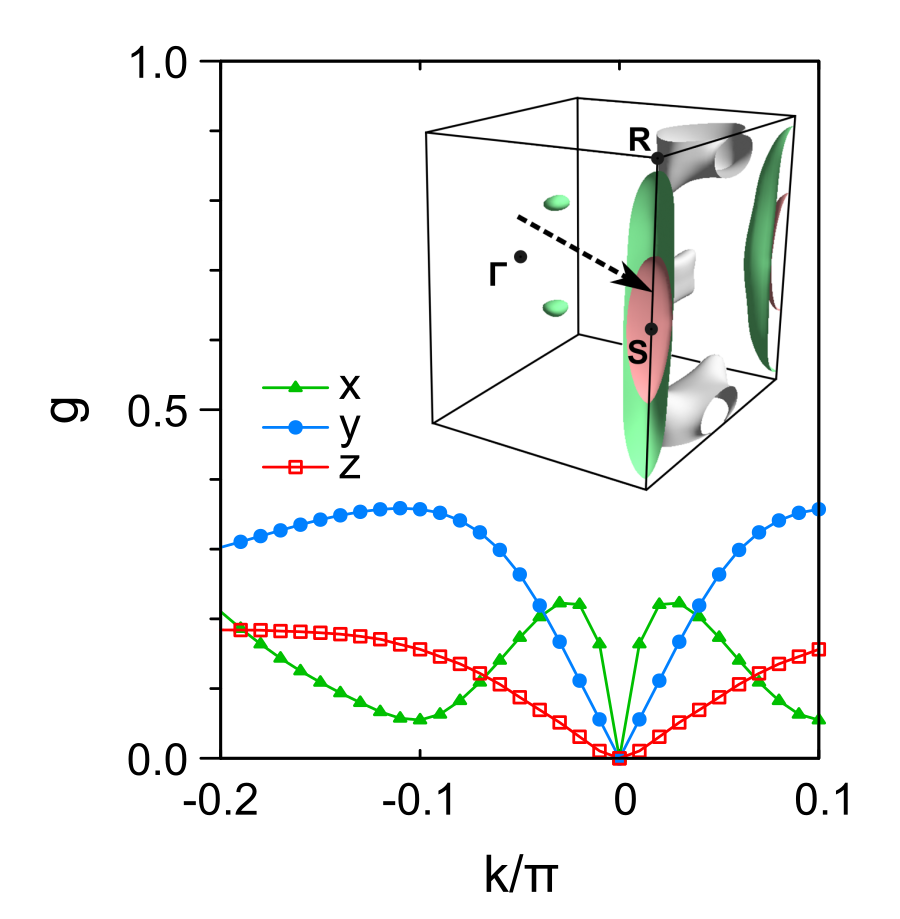}
		\caption{Intra-band $g$-factors in UCoGe calculated along the $k$ line $(\pi+k,\pi+k,0.2\pi)$. These g-factors quantify the energy splitting between Kramers' doublets for fields applied along the $x$, $y$, and $z$ directions. The value $g=1$ corresponds to usual spin-1/2. Inset: Fermi surface of UCoGe. The $g$ calculations are done along the arrow and for the bands producing the inner (pale red) Fermi surface. }
  \label{F_altermag}
	\end{figure}
\vglue 0.1 cm
\noindent {\it Vanishing $g$-factors: application to UCoGe:} The response of pseudospin to a  Zeeman field is generally described by the Hamiltonian $H_Z=\mu_B\sum_{i,j}g_{ij}H_i\tilde{\sigma}_j$. For spinless pseudospin, symmetry requires that $g_{i,j}=0$. This result has implications for pseudospin-singlet superconductors occurring on a Fermi surface surrounding the {\red above high symmetry lines}. In particular, since spinless pseudospin dominates the spin physics on such a Fermi surface, there is nearly no Pauli suppression of superconductivity for the field applied in {\it any} direction. This generalizes the well-known lack of Pauli suppression for in-plane fields observed in Ising superconductors \cite{Suh:2023,Wickramaratne:2023}.

One relevant material is UCoGe.  UCoGe has space group Pnma (SG 62) and exhibits critical fields that exceed the Pauli limit for all field directions \cite{Aoki:2019}. DFT shows that UCoGe exhibits quasi-2D cylindrical Fermi surfaces near the R-S line consistent with quantum oscillation measurements \cite{Bastien:2016}.  Along the R-S line, we predict spinless pseudospin. Figure 1 shows the DFT calculated $g$-factors near the R-S line and reveals that they all vanish on this line. Furthermore, they are small near this line, consistent with the large observed Pauli fields. 
UCoGe is often argued to be a pseudospin-triplet superconductor based on the large observed critical fields. Our results allow the possibility that this is a pseudospin-singlet superconductor and further necessitate a more careful analysis of spin fluctuations in such superconductors.

\vglue 0.1 cm
\noindent{\it Stoner spinless-pseudospin ferromagnets:}
Another important finding of our work is that spinless pseudospin induces Stoner's instabilities into novel magnetic states. To understand this, it is useful to revisit the Stoner instability of a single band into a ferromagnet.
Central to this is the local Coulomb repulsion on each lattice site $i$, $U n_{i,\uparrow}n_{i,\downarrow}$ (where $n_{i,s}$ gives the number of electrons with spin $s$). Treating this interaction within mean-field theory yields the usual Stoner criterion for a ferromagnetic instability, $UN(0)> 1$ where $N(0)$ is the density of states at the chemical potential.  The key observation for more general pseudospin is that the local Coulomb repulsion is statistical since it originates from the Pauli exclusion of the two degenerate spin states that comprise the band. Consequently, the same argument implies a Stoner instability into a pseudospin ferromagnet. When the pseudospin symmetry differs from that of usual spin-1/2, the resulting Stoner-driven magnetic state must have a different symmetry than a usual ferromagnet. 

{\blue Indeed, our earlier symmetry arguments imply that a Stoner-driven pseudospin ferromagnet will generally be non-collinear. To understand this, it is also worthwhile contrasting the Stoner instability of usual spin with that of pseudospin in materials with orthorhombic symmetry (similar arguments apply to cubic symmetry). For the usual spin, each spin component belongs to a different symmetry representation, hence symmetry requires a usual ferromagnetic transition into a uniaxial state. For spinless pseudospin, all pseudospin operators belong to the {\it same} symmetry representation, hence symmetry dictates that the resultant pseudospin ferromagnet is a linear combination of all three components. However, as Table 1 shows, each of these pseudospin components is related to real spin components through different momentum prefactors, implying that the spin quantization is strongly momentum-dependent and generally contains all three spin components.  In real space, this implies a generic non-collinear structure. }


{\blue We note that our general arguments imply that for a Fermi surface sufficiently close to the high symmetry line, we must always have a Stoner instability into a non-collinear magnetic state, independent of the details of the electronic structure. However, these symmetry arguments do not address what electronic properties determine the region of momentum space for which a pseudospin ferromagnet is the ground state.} {\blue To gain more insight}, we perform self-consistent one-loop (RPA) calculations for space group Pa$\overline{3}$. We consider a tight-binding model with four sites in a unit cell. We include nearest (with hopping parameter $t^\prime$), next-nearest neighbor hopping (with hopping parameter $t$), and nearest neighbor spin-orbit coupling $\lambda$ (see Methods). We study on-site repulsive Hubbard interaction $Un_{i\uparrow}n_{i\downarrow}$ with $i=1...4$. We compare all possible onsite magnetic orderings, which are captured by the 16 on-site particle-hole vertices $V_{i\uparrow\uparrow}$, $V_{i\downarrow\downarrow}$, $V_{i\uparrow\downarrow}$ and $V_{i\downarrow\uparrow}$. The corrections between these vertices are $\bf V=MV+V_0$, where ${\bf V}$ and ${\bf V_0}$ are vectors containing the 16 self-consistent and bare vertices, respectively.  $\bf M$ is a $16\times16$ matrix including all one-loop corrections. 

\begin{figure}[h]
	\centering
	\includegraphics[width=7cm]{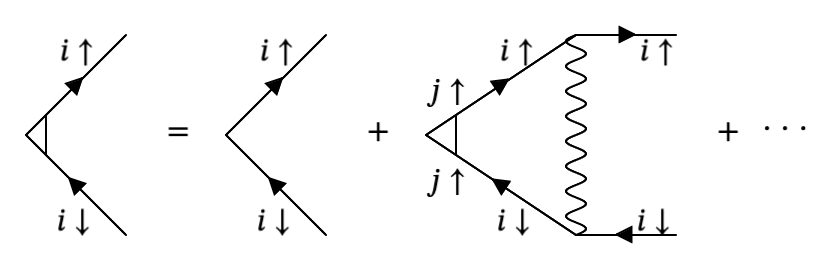}
	\includegraphics[width=7cm]{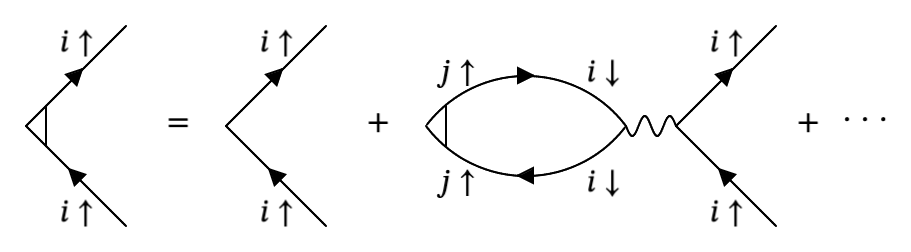}
	\caption{Two types of self-consistent one-loop Feynman diagrams, (Top) for output vertex with opposite spins, and (Bottom) for output vertex with the same spin. }
  \label{F:RPA_Feyn}
\end{figure}

{\red We only include intra-band particle-hole contribution while formulating} the calculation within the sublattice-spin basis, such that the matrix $\bf M$ involves two types of one-loop diagrams shown in Fig.\ref{F:RPA_Feyn}. The top type of diagram is relevant for output vertices with opposite spins. The particular correction shown here is from vertex $V_{j\uparrow\uparrow}$ to vertex $V_{i\uparrow\downarrow}$.  The bottom type of diagram is for output vertices with the same spins. The particular correction shown here is from vertex $V_{j\uparrow\uparrow}$ to vertex $V_{i\uparrow\uparrow}$. The corresponding matrix elements are written in Methods.

\begin{figure}[h]
	\centering
	\includegraphics[width=7cm]{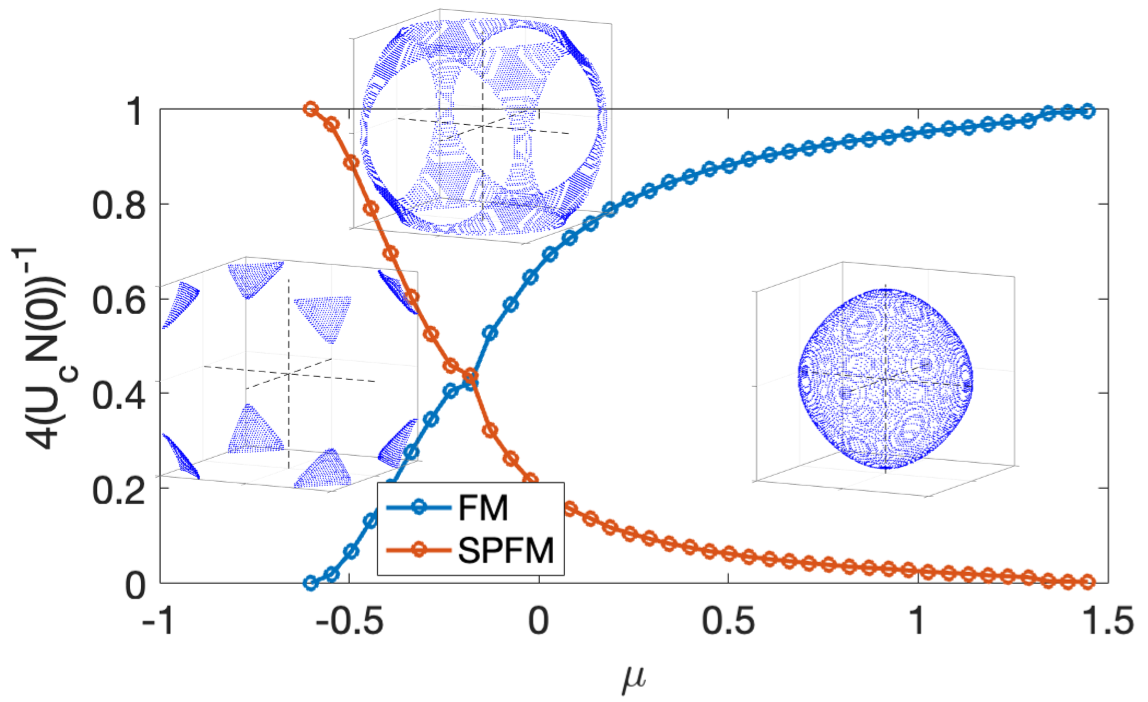}
	\caption{Inverse of critical interaction strength in Stoners' criteria, normalized by the density of states. As the Fermi surface evolves from a pocket near R-point $(\pi,\pi,\pi)$ to $\Gamma$-point, dominant magnetic instability changes from spinless pseudospin ferromagnetism (SPFM) to ferromagnetism (Fm). $t=0.15$, $t'=1$, $\lambda=-0.1$ are used. Fermi surfaces at $\mu=-0.4, 0,$ and $1.5$ are included.}
  \label{F:RPA_result}
\end{figure}

The eigenvalues of $\bf M$ have the form $\gamma UN(0)$, and Stoners' criteria is $\gamma U_cN(0)=1$. The dominant magnetic instability thus corresponds to the eigenvector of $\bf M$ with the largest $\gamma$. In Fig.\ref{F:RPA_result}, we show $\gamma=(U_cN(0))^{-1}$, for the ferromagnetic and spinless-pseudospin ferromagnetic states.
By increasing the chemical potential $\mu$, the Fermi surface evolves from a pocket near R-point $(\pi,\pi,\pi)$ to a pocket near $\Gamma$-point. For a Fermi surface near the $\Gamma$-point, the spinless feature is not present, and Stoners' mechanism gives a conventional ferromagnetic state. 

\begin{figure}[h]
		\centering
  \includegraphics[width=7cm]{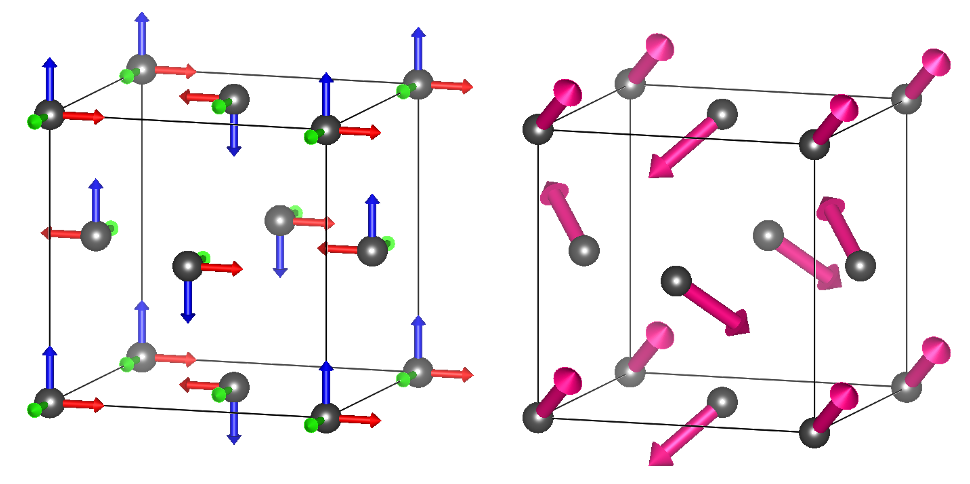}
		\caption{(Left) Staggered spin excitations with different orientations (denoted in different colors) are the spinless pseudospin excitations on the R-M ($k_x=k_y=\pi$) line of space group Pa$\overline{3}$ \cite{Momma:2011}. (Right) The three excitations share the same symmetry and generically mix into a non-collinear magnetic state. Similar states are found in space groups Pbca and Pnma. }
  \label{F:RPA_state}
\end{figure}

As the Fermi surface approaches the zone edges, spinless pseudospin dominates the low-energy excitations, and Stoners' mechanism gives rise to a {\blue non-collinear} magnetic state, as expected. {\red The exact transition point between a conventional ferromagnet and the spinless pseudospin ferromagnet depends on the ratio between SOC and nearest neighbor hopping: $\lambda/t'$, in Fig.~4 we use $|\lambda/t'|=0.1$. All spinless pseudospin representations break $T$ symmetry but preserve crystal symmetries as in Table 1. Moreover, as shown in Fig.\ref{F:RPA_state}, these excitations are staggered spin excitations, and generically mix into a non-collinear state.} 
The same states are found to be the leading instabilities in space group Pbca and Pnma, for Fermi surfaces near the R-S lines, and the resultant magnetic space groups are magnetic type 1 Federov groups for which all space group symmetries remain and only $T$ symmetry is broken.

Given that these different space groups all generate similar magnetic states from the same mechanism, it is reasonable to ask if they share any common physical properties. In the following, we discuss two such properties: $\pi/2$ spin gauge flux generated from the non-collinear magnetic order and symmetry required Weyl line nodes.


\vglue 0.1 cm
\noindent {\it Scalar spin chirality:} Non-collinear magnetic structures can give rise to scalar spin chirality  $\chi_{ijk}=\vec{S}_i\cdot(\vec{S}_j\times \vec{S}_k)$ which is closely linked to orbital magnetism and Berry curvature \cite{Wen:1989,Shindou:2001,Taguchi:2001,Machida:2010}. For all the magnetic groups we find here, $\chi_{ijk}=\pm 4m_xm_ym_z$ for any elementary triangle of nearest neighbor spins. This non-collinear magnetic structure implies spin-gauge flux for a spin-1/2 electron that travels each such triangle. We find this spin gauge flux is  $\pm \pi/2$ for all the spinless-pseudospin ferromagnetic states. Such spin-gauge fluxes are known to generate Berry curvature and an anomalous Hall effect (AHE). However,  
due to the presence of mirror symmetries in spinless-fermion ferromagnets, the total AHE will vanish. However, the application of a shear strain, which breaks a mirror symmetry, will allow both an orbital magnetization and an AHE to appear. {\blue The link between the magnetic order we find here for space group Pa$\overline{3}$  and orbital magnetization has been explored in fcc lattices \cite{Shindou:2001}. Our results generalize this to orthorhombic space groups and provide an electronic mechanism for this magnetic state.}


\vglue 0.1 cm
\noindent{\it Weyl lines:} We find that spinless-pseudospin ferromagnets generically have vanishing energy-band spin-splitting along lines in momentum space. These Weyl lines imply drumhead surface states on surfaces oriented along the mirror planes \cite{Chiu:2014,Chan:2016,Belopolski:2019}.
Near the $\Gamma$ point, symmetry requires that the Hamiltonian for spinless-pseudospin ferromagnets is  (a related Hamiltonian is discussed in Ref.~\onlinecite{Fernandes:2023})
\begin{equation}
H=\epsilon_0({\bm k})+\alpha_x k_zk_y\sigma_x+\alpha_y k_zk_x\sigma_y+\alpha_z k_xk_y \sigma_z
\end{equation}
where $\alpha_i$ are constants. This yields Weyl lines along the $x$, $y$, and $z$ axes that are each topologically protected by two of the three orthogonal mirror symmetries. On any mirror-symmetric loop around a Weyl line, the Berry phase is $\pi$ (here the relevant mirror symmetry is orthogonal to the Weyl line). 
This yields drumhead surface states on crystal surfaces that are mirror invariant \cite{Chan:2016}.  Key to the existence of these states is the position of possible additional Weyl lines away from the $\Gamma$ point. Such additional Weyl lines can imply that the drumhead surface states disappear.
Generically, using \cite{Elcoro:2021}, we find that such Weyl lines exist on the BZ boundary for all the spinless-pseudospin ferromagnets discussed here. The positions of all Weyl lines are shown in Fig. 3. This implies that drumhead surface states will appear on all $x$, $y$, and $z$ surfaces for spinless-pseudospin ferromagnets in space groups Pbca, Pbcn, Pnma, and Pa$\overline{3}$, and on the $y$ and $z$ surfaces for space group Pbcm.  In the SM we reveal these surface states for space group Pa$\overline{3}$. 


\begin{figure}[h]
	\centering
	\includegraphics[width=7cm]{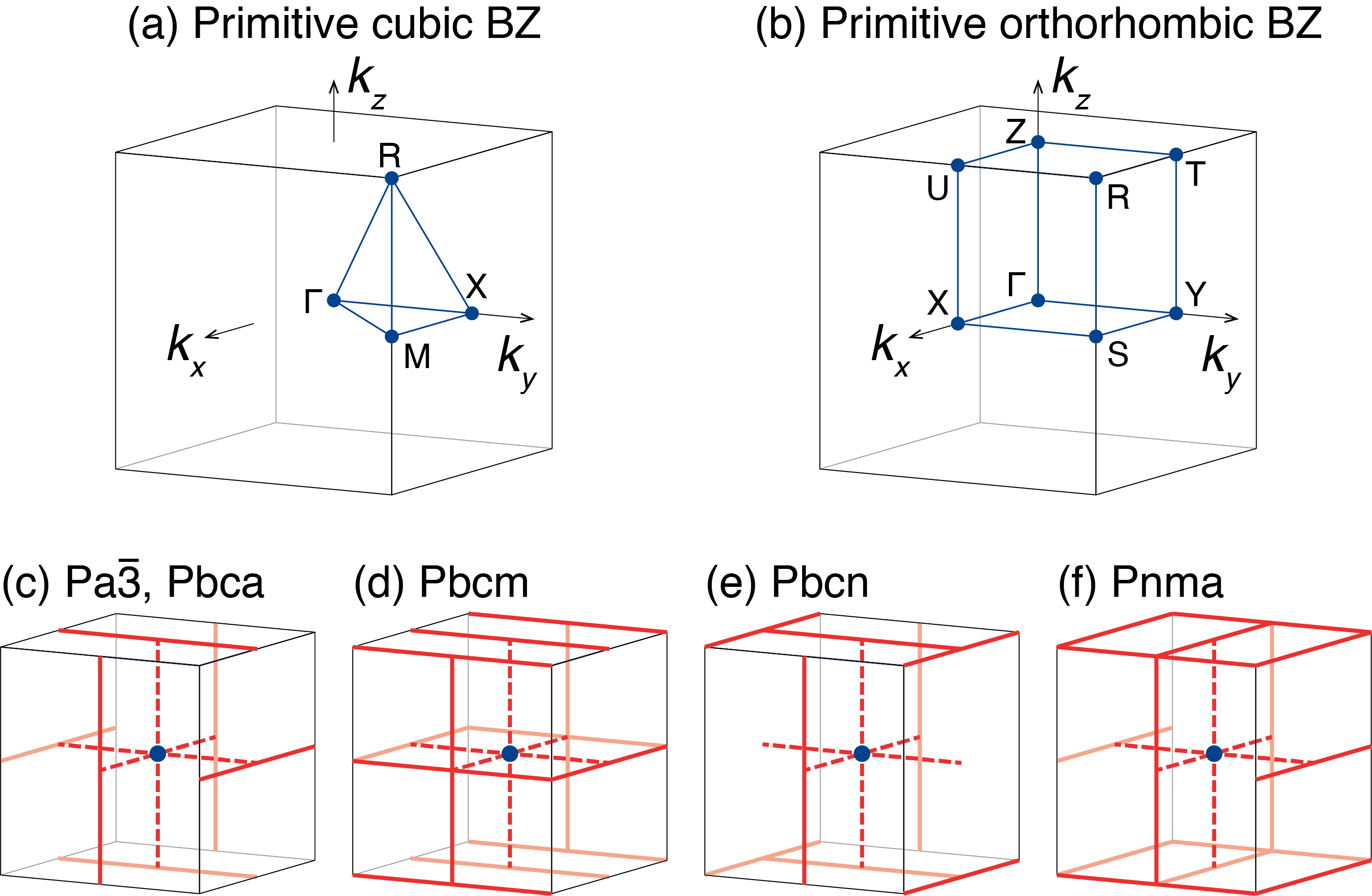}
		\caption{BZs for (a) primitive cubic and (b) primitive orthorhombic space groups. (c--f) Weyl fermion lines (in red) for the spinless-pseudospin ferromagnetic space groups are found here.}
\end{figure}
\vglue 0.1 cm
\noindent {\it Materials realizations of spinless pseudospin ferromagnets:} 
Using the  Magndata database \cite{Gallego:2016}  we highlight two material classes that reveal the magnetic states we find here. One is a group of 11 perovskite materials with space group Pnma. These have the form RTO$_3$, where R is a rare earth and T is a transition metal. DFT shows that many of these materials are metallic in the normal state (see for example \cite{Hasan:2022}).  The second is NiS$_{2-x}$Se$_x$ with space group Pa$\overline{3}$. {\red Under pressure and doping, the system goes through an insulator-metal transition \cite{Xu:2014,Friedemann:2016}. The translationally invariant non-collinear state has been reported in both phases under chemical pressure \cite{Yano:2016}. The metallic phase keeps a large density of states close to the insulator-metal transition \cite{Friedemann:2016}, suggesting the necessity for an itinerant mechanism to stabilize the non-collinear magnetic state.} DFT calculations on undoped NiS$_2$ show Fermi surfaces surrounding the R-point and additional Fermi surfaces surrounding the {\red high symmetry} lines \cite{Reiss:2022}, suggesting the applicability of our analysis. 


\vglue 0.1 cm
\noindent{\it Discussion and conclusions:} 

Our results emphasize Stoner instabilities of Fermi surfaces and thus emphasize the lowest energy degrees of freedom in stabilizing these novel magnetic states.  In principle, inter-band contributions can also generate such magnetic states \cite{Roig:2024}. In these cases, our results show that SOC plays an important role in further stabilizing these states.


In conclusion, we have identified a {\red symmetry property} of Bloch pseudospin that forbids it to couple to a Zeeman field on {\red certain high symmetry} lines in space groups Pbcm, Pbcn, Pbca, Pnma, and Pa$\overline{3}$. We have shown that this spinless {\red property} enables field-robust superconductivity and Stoner instabilities in non-collinear magnetic states that have no net magnetization. Finally, we have shown that these spinless-pseudospin ferromagnetic states generically have scalar spin chirality with $\pm \pi/2$ spin gauge flux and Weyl fermion lines that give rise to drumhead surface states. 

We thank Rafael M. Fernandes, Tomas Jungwirth, and Lian Li for useful discussions. D.F.A., T.S., M.W., and Y.Y were supported by the
National Science Foundation Grant No. DMREF 2323857.
S.S. was supported by JSPS KAKENHI Grants No.~JP23K03333 and No.~JP23K13056, and JST CREST Grant No. JPMJCR19T2.

\bibliography{refs}

\onecolumngrid
\section*{Methods}

For RPA calculation with a minimal tight-binding model. We considered $s$-orbitals on Wyckoff position 4a, given by 
 $(a_1,a_2,b_1,b_2)=\{(0,0,0),(0,1/2,1/2),(1/2,0,1/2),(1/2,1/2,0)\}$. We use the Pauli matrices $\tau_i$ and $\rho_i$ to differentiate between the $(a,b)$ and $(1,2)$ sublattices respectively, and the Pauli matrices $\sigma_i$ for the spin of these fermions.
The tight-binding Hamiltonian that includes all nearest-neighbor interactions is: 
\begin{equation}
		\begin{split}
H&=2t(\cos k_x+\cos k_y+\cos k_z)-\mu\\&+t'\left(\cos\frac{k_x}{2}\cos\frac{k_z}{2}\tau_x+\cos\frac{k_y}{2}\cos\frac{k_z}{2}\rho_x+\cos\frac{k_x}{2}\cos\frac{k_y}{2}\tau_x\rho_x\right)+H_{SOC1}+H_{SOC2}
\\H_{SOC1}&=\lambda\left(\sin\frac{k_x}{2}\sin\frac{k_y}{2}\tau_y\rho_x\sigma_x+\sin\frac{k_y}{2}\sin\frac{k_z}{2}\rho_y\sigma_y+\sin\frac{k_x}{2}\sin\frac{k_z}{2}\tau_y\rho_z\sigma_z\right)\\
H_{SOC2}&=a\left(\cos\frac{k_y}{2}\cos\frac{k_z}{2}\rho_y\sigma_z+\cos\frac{k_y}{2}\cos\frac{k_z}{2}\tau_z\rho_y\sigma_x+\cos\frac{k_x}{2}\cos\frac{k_z}{2}\tau_y\sigma_y\right.\\&\left.+\cos\frac{k_x}{2}\cos\frac{k_z}{2}\tau_y\rho_z\sigma_x+\cos\frac{k_x}{2}\cos\frac{k_y}{2}\tau_x\rho_y\sigma_z+\cos\frac{k_x}{2}\cos\frac{k_y}{2}\tau_y\rho_x\sigma_y\right)
\end{split}
	\end{equation}

{\red RPA calculation on Pa$\overline{3}$ studies the doubly generate band with the highest energy. The top panel of Fig.\ref{F:RPA_Feyn} gives matrix element $M_{(i\uparrow,\downarrow),(j\uparrow,\uparrow)}={UN(0)}\left\langle\sum_{a,b=I}^{II}u^a_{j\uparrow}u^{a*}_{i\uparrow}u^{b}_{i\downarrow}u^{b*}_{j\uparrow}\right\rangle$. Here, $u^I$ and $u^{II}$ are the two degenerate eigenvectors of the Hamiltonian, and $u^a_{j\uparrow}$ is the $j\uparrow$ element of the eigenvector $u^a$. The average $\langle ... \rangle$ is over all states on the Fermi surface. $N(0)$ is the density of states at the Fermi level.  The bottom panel gives the matrix element
$M_{(i\uparrow,\uparrow),(j\uparrow,\uparrow)}=-{UN(0)}\left\langle\sum_{a,b=I}^{II}u^a_{j\uparrow}u^{a*}_{i\downarrow}u^{b}_{i\downarrow}u^{b*}_{j\uparrow}\right\rangle$. 

Mirror symmetries are $M_x\propto\tau_y\rho_x\sigma_x$, $M_y\propto\rho_y\sigma_y$, and $M_z\propto\tau_y\rho_z\sigma_z$. Local staggered spin excitations $\tau_z\rho_z\sigma_x$, $\tau_z\sigma_y$, and $\rho_z\sigma_z$ preserve all mirror symmetries, and are pseudospin excitations on the zone edge. These three states share the same symmetry as $k_xk_y\sigma_z$, so they generically mix, giving rise to non-collinear states. The exact mixing ratio depends on higher order terms in the theory, e.g. $H_{SOC2}$. }

\clearpage

\section*{Supplementary Information}

\subsection{Symmetry arguments for spinless pseudospin}
Here, we will first review the symmetry arguments for the double nodal line \cite{Takahashi:2017,Li:2021} in the absence of SOC. We then include SOC to reveal how the nodal eight-fold degeneracy is broken into four Kramers' doublets. 

The eight-fold band degeneracy requires four-fold orbital degeneracy. Consider the symmetry elements that preserve a momentum point on the nodal line $(\pi,\pi,k_z)$. They are generated by $\{E,\widetilde{M_x},\widetilde{M_y},IT\}$, where $\widetilde{M_{x}}$ is the usual mirror symmetry $x\rightarrow-x$ followed by a translation ${\bf {t_{x}}}=(t_{x}^x,t_{x}^y,t_{x}^z)$. The origin is chosen to be the inversion center. We have $\widetilde{M_{x}}^2=\{0,2t^y_x,2t^z_x\}$, $\widetilde{M_{x}}I=\{2t^x_x,2t^y_x,2t^z_x\}I\widetilde{M_{x}}$, and $I^2=T^2=1$. Here $\{...\}$ denotes translation. The spin is ignored in the absence of SOC. We show that the four states (below labeled as $1 \rightarrow 4$):
	\begin{equation}
		|\psi\rangle, \ \widetilde{M_x}IT|\psi\rangle, \ \widetilde{M_x}\widetilde{M_y}|\psi\rangle, \ \widetilde{M_y}IT|\psi\rangle
	\end{equation}
 are degenerate orthogonal eigenstates. 
They follow from two two-fold Kramers' degeneracies. The first Kramers' degeneracy arises from $\widetilde{M_x}IT$. This satisfies $(\widetilde{M_x}IT)^2=\{2t^x_x,0,0\}=\exp(2ik_xt_x^{x})=-1$, (here we used the fact that the $T$ commutes with other symmetry operations).  This symmetry makes states (1,2) as well as (3,4) orthogonal to each other. The second Kramers' degeneracy arises from $\widetilde{M_y}IT$, which satisfies $(\widetilde{M_y}IT)^2=\exp(2ik_yt_y^{y})=-1$. This makes states (1,4) as well as (2,3) orthogonal to each other.
For these two Kramers' degeneracies to be independent, states (1,3): $|\psi\rangle$ and $\widetilde{M_x}\widetilde{M_y}|\psi\rangle$ need to be orthogonal. This requires $\{\widetilde{M_x},\widetilde{M_y}\}=0$, such that $|\psi\rangle$ can be chosen as an eigenstate of $\widetilde{M_x}$, while $\widetilde{M_x}\widetilde{M_y}|\psi\rangle$ is an orthogonal eigenstate with the opposite $\widetilde{M_x}$ eigenvalue. In summary, without SOC, the eight-fold band degeneracy requires $\exp(2ik_xt_x^{x})=\exp(2ik_yt_y^{y})=-1$ and $\{\widetilde{M_x},\widetilde{M_y}\}=0$. The former condition gives  $t_x^{x}=t_y^{y}=1/2$ and the latter condition restricts $(t_x^{y},t_y^{x})=(0,1/2)$ or $(1/2,0)$. 

We now include the SOC. The time reversal operation now satisfies $T^2=-1$. The bands have the usual Kramers' degeneracy from $IT$. The mirror operators $\widetilde{M_{x,y}}$ now include the spin part $i\sigma_{x,y}$. This leads to $[\widetilde{M_x},\widetilde{M_y}]=0$. There is now no additional degeneracy arising from other symmetry operators involving $T$, since $\widetilde{M_x}$ and $\widetilde{M_y}$ commute with the Hamiltonian and with each other, so define four non-degenerate orthogonal eigenstates. Hence the eight-fold degeneracy splits into four Kramers' pairs. 
 
We now prove that each Kramers' pair shares the same $\widetilde{M_x}$ eigenvalues. The proof for $\widetilde{M_y}$ is similar. Since $\widetilde{M_x}^2=(i\sigma_x)^2\{0,2t^y_x,2t^z_x\}=-\exp(2ik_yt_x^y+2ik_zt_x^z)$, the eigenvalue of $\widetilde{M_x}$ is $\pm i\exp(ik_yt_x^y+ik_zt_x^z)$. 
Let $|\psi\rangle$ be an eigenstate of $\widetilde{M_x}$, then consider its Kramers' partner $TI|\psi\rangle$:
\begin{equation}
\begin{split}
&\widetilde{M_x}|\psi\rangle=ie^{ik_yt_x^y+ik_zt_x^z}|\psi\rangle\\
\Rightarrow&\widetilde{M_x}TI|\psi\rangle=\{2t_x^x,2t_x^y,2t_x^z\}TI\widetilde{M_x}|\psi\rangle=e^{2ik_xt_x^x+2ik_yt_x^y+2ik_zt_x^z}TI(ie^{ik_yt_x^y+ik_zt_x^z}|\psi\rangle)\\=&(-e^{2ik_xt_x^x})ie^{ik_yt_x^y+ik_zt_x^z}TI|\psi\rangle=ie^{ik_yt_x^y+ik_zt_x^z}TI|\psi\rangle
\end{split}
\end{equation}
Hence $TI|\psi\rangle$ is another eigenstate with the same $\widetilde{M_x}$ eigenvalue as $|\psi\rangle$. Since the two Kramers' partners share the same mirror eigenvalues, Pauli matrices made from these Kramers' partners must be invariant under the two mirror operations $\widetilde{M_x}$ and $\widetilde{M_y}$ \cite{Cavanagh:2022,Suh:2023}. This implies that all the Kramers' doublets form spinless pseudospin.

\subsection{Tight-binding Hamiltonian for other space groups}
For space groups Pbca, Pnma, and Pa$\overline{3}$ we considered $s$-orbitals on Wyckoff position 4a. For space groups Pbca and Pa$\overline{3}$ these are given by 
 $(a_1,a_2,b_1,b_2)=\{(0,0,0),(0,1/2,1/2),(1/2,0,1/2),(1/2,1/2,0)\}$. We use the Pauli matrices $\tau_i$ and $\rho_i$ to differentiate between the $(a,b)$ and $(1,2)$ sublattices respectively, and the Pauli matrices $\sigma_i$ for the spin of these fermions.
The tight-binding Hamiltonian that includes all nearest-neighbor interactions is: 
\begin{equation}
  \label{eq:tight-binding_Pbca}
		\begin{split}
H&=2t_x\cos k_x+2t_y\cos k_y+2t_z\cos k_z-\mu\\&+t_{ab}\cos\frac{k_x}{2}\cos\frac{k_z}{2}\tau_x+t_{12}\cos\frac{k_y}{2}\cos\frac{k_z}{2}\rho_x+t_{a1b2}\cos\frac{k_x}{2}\cos\frac{k_y}{2}\tau_x\rho_x+H_{SOC1}+H_{SOC2}
\\H_{SOC1}&=\lambda_1\sin\frac{k_x}{2}\sin\frac{k_y}{2}\tau_y\rho_x\sigma_x+\lambda_2\sin\frac{k_y}{2}\sin\frac{k_z}{2}\rho_y\sigma_y+\lambda_3\sin\frac{k_x}{2}\sin\frac{k_z}{2}\tau_y\rho_z\sigma_z\\
H_{SOC2}&=a_1\cos\frac{k_y}{2}\cos\frac{k_z}{2}\rho_y\sigma_z+a_2\cos\frac{k_y}{2}\cos\frac{k_z}{2}\tau_z\rho_y\sigma_x+a_3\cos\frac{k_x}{2}\cos\frac{k_z}{2}\tau_y\sigma_y\\&+a_4\cos\frac{k_x}{2}\cos\frac{k_z}{2}\tau_y\rho_z\sigma_x+a_5\cos\frac{k_x}{2}\cos\frac{k_y}{2}\tau_x\rho_y\sigma_z+a_6\cos\frac{k_x}{2}\cos\frac{k_y}{2}\tau_y\rho_x\sigma_y
\end{split}
	\end{equation}
For space group Pbca, the coefficients are generically different. For space group Pa$\overline{3}$, the following constraints are imposed by symmetry:
\begin{equation}
 \label{eq:tight-binding_Pa3_constraint}
 t_x = t_y = t_z =: t, \quad
 t_{ab} = t_{12} = t_{a1b2} =: t', \quad
 \lambda_1 = \lambda_2 = \lambda_3 =: \lambda, \quad
 a_1 = a_4 = a_6 =: a, \quad
 a_2 = a_3 = a_5 =: a'.
\end{equation}

For space group Pnma, we denote the 4a Wyckoff positions as $(a_1,a_2,b_1,b_2)=\{(0,0,0),(1/2,0,1/2),(0,1/2,0),(1/2,1/2,1/2)\}$.
The tight-binding Hamiltonian is: 
	\begin{equation}
		\begin{split}
H&=2t_x\cos k_x+2t_y\cos k_y+2t_z\cos k_z-\mu\\&+t_{ab}\cos\frac{k_y}{2}\tau_x+t_{12}\cos\frac{k_x}{2}\cos\frac{k_z}{2}\rho_x+t_{a1b2}\cos\frac{k_x}{2}\cos\frac{k_y}{2}\cos\frac{k_z}{2}\tau_x\rho_x+H_{SOC1}+H_{SOC2}
\\H_{SOC1}&=\lambda_1\sin\frac{k_x}{2}\sin\frac{k_y}{2}\cos\frac{k_z}{2}\tau_x\rho_y\sigma_x+\lambda_2\sin\frac{k_y}{2}\sin{k_z}\tau_y\sigma_y+\lambda_3\sin\frac{k_x}{2}\sin\frac{k_z}{2}\tau_z\rho_y\sigma_z\\
H_{SOC2}&=a_1\cos\frac{k_x}{2}\cos\frac{k_z}{2}\tau_0\rho_y\sigma_y+a_2\cos\frac{k_x}{2}\cos\frac{k_z}{2}\tau_z\rho_y\sigma_x+a_3\cos\frac{k_y}{2}\tau_y\rho_0\sigma_z\\&+a_4\cos\frac{k_y}{2}\tau_y\rho_z\sigma_x+a_5\cos\frac{k_x}{2}\cos\frac{k_y}{2}\cos\frac{k_z}{2}\tau_x\rho_y\sigma_y+a_6\cos\frac{k_x}{2}\cos\frac{k_y}{2}\cos\frac{k_z}{2}\tau_y\rho_x\sigma_z\\&+a_7\cos\frac{k_y}{2}\sin\frac{k_x}{2}\sin\frac{k_z}{2}\tau_y\rho_x\sigma_x+a_8\cos\frac{k_x}{2}\sin\frac{k_y}{2}\sin\frac{k_z}{2}\tau_x\rho_y\sigma_z+a_9\cos\frac{k_x}{2}\sin\frac{k_y}{2}\sin\frac{k_z}{2}\tau_y\rho_x\sigma_y
\end{split}
	\end{equation}

\subsection{Drumhead surface states}
In this section, we demonstrate the drumhead surface states for space group Pa$\overline{3}$.
For this purpose, we use the tight-binding model in Eq.~\eqref{eq:tight-binding_Pbca} under the conditions \eqref{eq:tight-binding_Pa3_constraint}.
Furthermore, additional SOC terms allowed in the space group symmetry are considered here:
\begin{align}
 H_{\text{SOC4}} &= \lambda (\delta_1 \cos k_x + \delta_2 \cos k_y + \delta_3 \cos k_z) \sin\frac{k_x}{2} \sin\frac{k_y}{2} \tau_y \rho_x \sigma_x \notag \\
 & + \lambda (\delta_3 \cos k_x + \delta_1 \cos k_y + \delta_2 \cos k_z) \sin\frac{k_y}{2} \sin\frac{k_z}{2} \rho_y \sigma_y \notag \\
 & + \lambda (\delta_2 \cos k_x + \delta_3 \cos k_y + \delta_1 \cos k_z) \sin\frac{k_x}{2} \sin\frac{k_z}{2} \tau_y \rho_z \sigma_z,
\end{align}
which lift accidental degeneracies of the energy bands and enable us to see the existence of the surface states.

\begin{figure}[h]
		\centering
		\includegraphics[width=10cm]{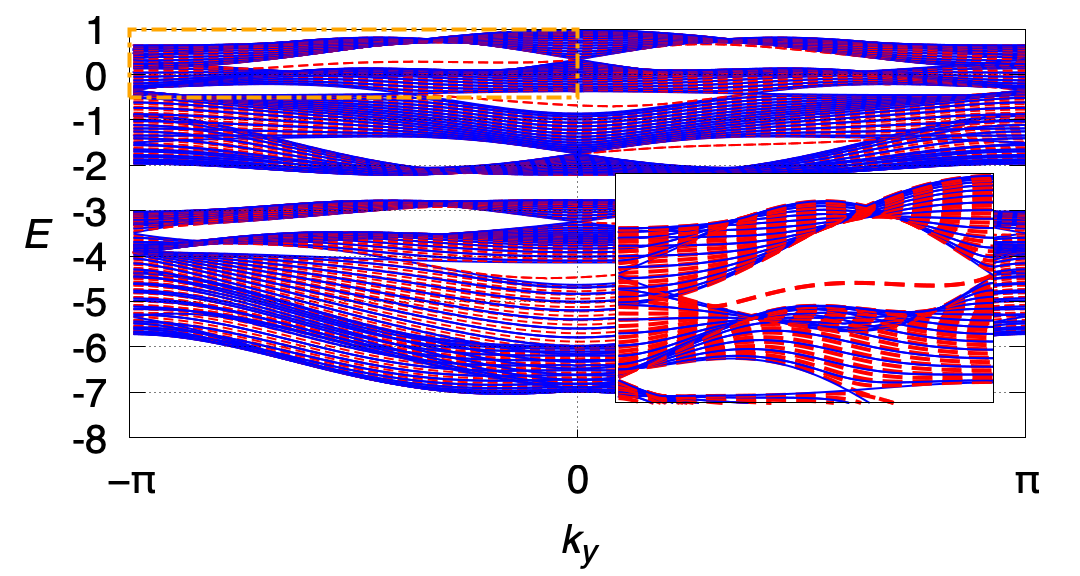}
		\caption{Weyl-line node driven topological surface states. Parameters are set to $(t, t', \lambda, a, a', \delta_1, \delta_2, \delta_3, h, \mu) = (-0.2, 0.2, 1, 1, -1, 0.2, 0.5, 0.5, 1, 2.5)$. The blue solid and red dashed lines represent the spectrum under the PBC and OBC along the $z$ axis, respectively. The inset indicates the enlarged energy spectrum in the orange dash-dotted box.}
  \label{drumhead}
\end{figure}

To discuss the altermagnetic-like state induced by the Stoner instability for spinless pseudospin, we treat the magnetic order as a molecular field: $h (\rho_z \sigma_z + \tau_z \rho_z \sigma_x + \tau_z\sigma_y)$.
Figure~\ref{drumhead} shows energy spectra on the slice with $k_x = \pi/2$.
The blue solid lines are obtained by considering periodic boundary conditions (PBCs) for all directions, whereas the red dashed lines represent the spectra under an open boundary condition (OBC) along the $z$ axis and PBCs along the other $x$ and $y$ directions.
There are several surface bands in the bulk gap, which originate from the topology of the Weyl lines in Fig.~3(c) of the main text.
For example, the drumhead surface states emerging in the $-\pi < k_y < 0$ region of the upper bands (the inset of Fig.~\ref{drumhead}) are characterized by the Zak phase,
\begin{equation}
 \gamma_n(k_x, k_y) = \frac{1}{i} \int_{-\pi}^{\pi} \mathrm{d}k_z \langle u_n(\bm{k}) | \partial_{k_z} | u_n(\bm{k}) \rangle \pmod{2\pi},
\end{equation}
which is quantized into a $Z_2$ variable by the mirror symmetry $\widetilde{M}_z$.
Here $n$ represents a band index.
We calculate the Zak phase $\gamma_n(k_x, k_y)$ by choosing $n$ as the top band, and confirm that it takes a nontrivial value $\pi$ for $-\pi < k_y < 0$, while it is zero for $0 < k_y < \pi$.
These results correspond to the presence or absence of drumhead surface states.

 \end{document}